# Impact of Spatial Separation of Type-II GaSb Quantum Dots from the Depletion Region on the Conversion Efficiency Limit of GaAs Solar Cells

A. Kechiantz[1, 3, *], A. Afanasev[1], J.-L. Lazzari[2]

[1]Department of Physics, The George Washington University, 725 21st Street, NW, Washington, DC, 20052, USA; kechiantz@gwu.edu; afanas@gwu.edu
[2]Centre Interdisciplinaire de Nanoscience de Marseille, CINaM, UMR 7325 CNRS – Aix-Marseille Université, Campus de Luminy, Case 913, 163 Avenue de Luminy, 13288 Marseille Cedex 9, France; lazzari@cinam.univ-mrs.fr
[3]On leave of absence: Institute of Radiophysics and Electronics, National Academy of Sciences, 1 Brothers Alikhanyan st., Ashtarak 0203, Armenia; mrs_armenia@yahoo.com

**Abstract**
The purpose of this work is to look for a practical structure for application of quantum dots (QD) in solar cells in order to enhance sub-band gap photon absorption. We focus on a stack of strain-compensated GaSb/GaAs type-II QDs. We propose a novel structure with GaSb/GaAs type-II QD absorber embedded in the p-doped region of ideal solar cell, but spatially separated from the depletion region. We developed the model and used the detailed balance principle along with Poisson and continuity equations for calculating of the energy band bending along with the photocurrent and the dark current, and the conversion efficiency of the cell. Our model takes into account both single-photon and double-photon absorption as well as non-radiative processes in QDs and predicts that oncentration from 1-sun to 500-sun raises the efficiency from 30% to 50%. We showed that accumulation of charge in the QD absorber is the clue to understanding of potentially superior performance of the proposed solar cell. An attractive feature of the proposed solar cell is that QDs do not reduce the open circuit voltage but facilitate generation of the additional photocurrent to the extent that photovoltaic characteristics reduce to that of ideal IB solar cell while the efficiency meets the Luque-Marti limit. It should be noted that, although non-radiative processes like relaxation in QDs and recombination through QDs degrade photovoltaic characteristics of the proposed solar cell, its conversion efficiency is still predicted to be above the Shockley-Queisser limit by 5% to 10%. This study is an important step toward producing practical solar cells that benefit from additional photocurrent generated by sub-band gap photons.

**\*Correspondence**
A. Kechiantz, Dept of Physics, The George Washington University, Washington, DC, USA. E-mail: kechiantz@gwu.edu;

# 1. Introduction

A survey of the energy domain shows a sustainable progress in development of solar cells and their diversification for last decades. This progress is firmly connected with progress in nanotechnology and new ideas and physical effects used in photovoltaics [1].

One of such ideas was exploitation of low-dimensional structures such as quantum wells [2] and quantum dots (QDs) [3] for generation of additional photocurrent by low energy photons, in particular, due to two-photon absorption [4]. The latter is in fact a non-linear optical effect [5]. For enforcing this two-photon absorption, A. Luque and A. Marti offered to use a resonant absorption via intermediate band (IB) electronic states embedded into the absorber [6]. As a non-linear effect, two-photon absorption benefits from concentration of incoming sunlight. The conversion efficiency limit increases up to 63% for such IB solar cells. Both QD and Luque-Marti concepts attracted much attention and fuelled a search of new structures and physical principles able to push up the conversion efficiency of solar cells above the 37% of Shockley-Queisser limit.

Practical realization of these concepts encountered technological problems connected with formation of high-density QD arrays with low defect densities in solar cells. For instance, tens-nm thick GaAs spacers between QD arrays can be used for achieving of low defect densities [7], however, at expense of carrier extraction from QDs [8]. A few nm thin spacers are good for electron tunneling, however, lattice-relaxation facilitates formation of misfit dislocations and generation of deep-level defects between QDs in such stack of high-density QD arrays [9]. Photon absorption curves with Urbach tails extending far below the semiconductor band gap disclosed presence of deep level defects in the stack of QDs [8]. Another option was formation of strained QDs at low deposition temperature. However, such strain generated intrinsic defects at InAs/GaAs interfaces and was accumulated with increasing the number of QD arrays in the stack [10].

An essential progress has been achieved recently by incorporating of strain-compensated QD arrays into solar cells. A 5% increase of the photocurrent by InAs/GaAs type-I QDs and 17.8% conversion efficiency was measured in such QD GaAs solar cells [11]. Even more impressive is that effective management and proper balancing of the strain in the stack of 40 InAs/GaAs type-I QD layers enabled to achieve 0.5% absolute efficiency improving in comparison with the control single-junction GaAs cell [12]. Such success permitted the same group to challenge an ability of transferring the QD solar cell technology to large-area double-junction and triple-junction solar cells [13].

Another important development in the field was recognition [5, 14] that type-II QDs show potential for solar cell applications. There are also relevant developments with indirect band gap Si/Ge type-II QDs [15, 16].

The next step towards practical realization of QD and Luque-Marti concepts will be search of semiconductor structures reducing the dark current associated with recombination in the depletion region of p-n-junction [5]. Experiments shown that QDs embedded in the depletion region facilitated generation of both additional photocurrent and additional dark current [4]. For achieving efficiency above the Shockley-Queisser limit, the depletion region should be the high crystalline quality while QDs would not increase the dark current too much [15]. However, very often, electron-hole recombination in QDs raised additional dark current to the extent that it reduced open circuit voltage and kept the conversion efficiency of QD solar cells far below the Shockley-Queisser limit [17, 18].

The IB experiments with QDs reflect the fact that the dark current of solar cells, $j_{Dark}$, is related to recombination rate in the depletion region. When a bias $V$ is applied, this rate raises as $G_T(np/n_i^2 - 1)$, where $n_i$ is the intrinsic density of carriers; $n$ and $p$ are the density of injected electrons and holes, respectively, $np/n_i^2 \approx exp(eV/kT)$; $G_T$ is the rate of thermal generation of carriers, $G_T = \gamma_R n_i^2$; and $\gamma_R$ is the recombination coefficient related to the effective cross-section of electron-hole capture [19].

There are three reasons why location of QDs within the depletion region increases the dark current. First, the built-in field of depletion region facilitates the electron tunneling into and from the QD confined electronic states. Such tunneling may essentially increase the dark current of solar cells because thermal generation of carriers is most intensive in the narrow band gap regions, $G_T \sim \gamma_R exp(-\varepsilon_g/kT)$, in particular, in QDs where the band gap, $\varepsilon_g$, is smaller than in the rest of host semiconductor.

The next reason is that the effective cross-section $\gamma_R$ of recombination is high in the type-I QDs since such QDs capture both electrons and holes. It should be noted that the effective cross-section $\gamma_R$ in the type-II QDs is smaller than that in type-I QDs because type-II QDs spatially separate mobile electrons from confined holes.

The third reason is that embedding of QDs in the depletion region may give rise of additional defects in this technologically very sensitive part of solar cells.

Let us recall that a quick response is often required for optoelectronic applications [20]. The built-in field quickly separates generated electrons from holes in the depletion region of p-n-junction. Therefore, the depletion region became a "default" location for absorber of incoming photons in devices that generated a photocurrent in response to incoming photons, for instance, in photo-detectors. As a "default" location of QDs, the depletion region is routinely used also in solar cells. However, it is just a habit due to an assumption that quick spatial separation of electrons from holes would increase the conversion efficiency of solar cells.

Looking for a new practical structure for QD solar cell application, we have focused on a stack of strain-compensated GaSb/GaAs type-II QDs and application of such stack as the QD absorber of sub-band gap photons [5].

In this paper we will show that removing of the QD absorber from the "default" location may solve the QD problem for solar cells. The key idea of this solution is that QD absorber spatially separated from the depletion region is unable to facilitate generation of the additional dark current in the depletion region. We proposed a new GaSb/GaAs QD solar cell, where the QD absorber is embedded in the p-doped region of the GaAs p-n-junction but spatially separated from the depletion region as shown in Figure 1.

We developed a model and used the detailed balance principle along with Poisson and continuity equations for calculating of the energy band bending, the photo and dark currents, and the conversion efficiency of proposed QD solar cell. Our model takes into account both single- and two-photon absorption as well as non-radiative processes, like intra-band relaxation of holes in QDs and inter-band electron-hole recombination in QDs, both associated with QDs embedded in the otherwise ideal solar cell.

## 2. Theory

*2.1. Spatially separated type-II QD absorber*

Figure 1 illustrates the structure of proposed GaAs QD IB solar cell. A thin $p^+$-doped $Al_xGa_{1-x}As$ buffer layer grown on an $n^+$-doped GaAs substrate separates the QD absorber from the $p^+$-doped $Al_xGa_{1-x}As$ cap layer. This QD absorber is a stack of GaSb strained type-II QD layers alternating with p-doped $Al_xGa_{1-x}As$ spacers. All layers of the stack are within the electron diffusion length distance from the depletion region.

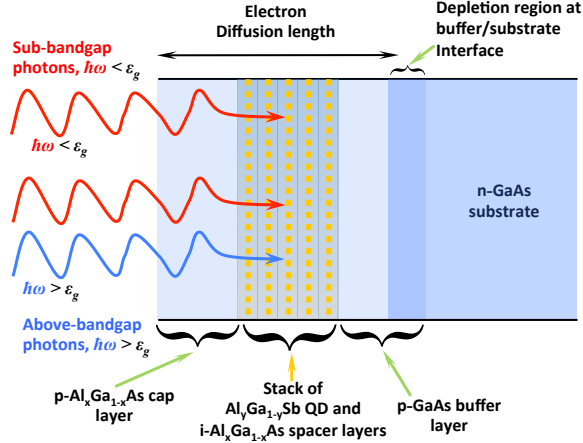

**Figure 1.** A simplified layout of GaAs solar cell with spatially separated GaSb/GaAs type-II QD absorber. The QD absorber is a stack of GaSb type-II QD layers alternating with $Al_xGa_{1-x}As$ spacers. The GaAs buffer layer spatially separates the QD absorber from the depletion region of GaAs p-n-junction.

We assume that the $Al_xGa_{1-x}As$ spacers are thick enough for we can ignore the electron tunneling of confined holes from and between QDs in the valence band. We also assume that the p-doped buffer forms an ideal p-n-junction with the n-doped substrate such that it does not allow electron tunneling through the depletion region and from the n-doped substrate into the electronic states confined in QDs.

*2.2. Blocking barrier $\varepsilon_B$*

Figure 2 displays the energy band diagram of proposed solar cell. There is a blocking barrier $\varepsilon_B$ in the conduction band that spatially separates mobile electrons of the QD absorber from the depletion region. Poisson equation determines $\varepsilon_B = \psi_B - \psi_A$ value of this barrier

$$\frac{\partial^2 \psi}{\partial x^2} = \frac{2kT}{L_{Deb}^2}\left(1 - \frac{n_B}{N_A}exp\frac{\psi_B - \psi}{kT} + \frac{N_B}{N_A}exp\frac{\psi - \psi_B}{kT}\right) \qquad (1)$$

where $\psi(x)$ is the bending of the conduction band edge, which equals $\psi_A$ in the absorber and $\psi_B$ in the buffer, $\psi_B = kT \times ln[n_i^2/(N_B N_n)]$; $N_A$, $N_B$ and $N_n$ are the doping of absorber, buffer and n-doped substrate, respectively; $n_B$ is the electron density in the buffer; $L_{Deb}$ is the Debye length in the absorber, $L_{Deb} = \sqrt{2\varepsilon\varepsilon_0 kT/e^2 N_A}$; and $n_i$ is the intrinsic density of carriers in the buffer. The boundary conditions for Equation (1) are: $\partial\psi/\partial x = 0$ and $\psi = \psi_A$ in the middle, $x = L_A/2$, of absorber; $\partial\psi/\partial x = 0$ and $\psi = \psi_B$ in the buffer where the electron density is $n_B$.

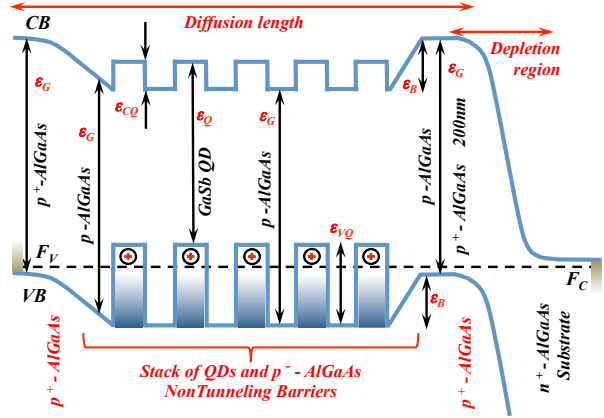

**Figure 2.** Energy band diagram of GaAs solar cell with GaSb/GaAs type-II QD absorber spatially separated from the depletion region of p-n-junction.

Parameter $n_B$ of Equation (1) is a function of bias voltage, $V$, and the flux, $G$, of solar photons. For instance, $n_B = n_{B0} exp(eV/kT)$ when $V$ bias is applied in the dark, $G = 0$, but $n_B = n_{B0}$ and $\varepsilon_B = \varepsilon_{B0}$ when $G = V = 0$, where $n_{B0} \equiv (n_i^2/N_B)$.

*2.3. Electron transfer into the conduction band*

Let us assign indexes $A$ to the QD absorber, $B$ to the buffer layer, $C$ to the conduction band, $N$ to the non-radiative electron transitions, $Q$ to the type-II QDs, and $V$ to the valence band. Then $G_{CV}$ is the rate of radiative electron transitions from the valence band into the conduction band due to photon absorption while $R_{NV}$ is the rate of non-radiative intra-band relaxation of holes from mobile into confined electronic states of QDs in the valence band of QD absorber.

When QDs facilitate the two-photon absorption, both conventional $GR_C$ and additional $GR_Q$ mechanisms transfer electrons into the conduction band of QD absorber,

$$GR = G_{CV} - R_{CV}\left(exp\frac{\mu_C}{kT} - 1\right) + G_{CQ} - (R_{CQ} + R_{NC})\left(exp\frac{\mu_C - \mu_Q}{kT} - 1\right) \quad (2)$$

where $GR$ is the net rate of electron transfers into the conduction band,

$$GR_C = G_{CV} - R_{CV}\left(exp\frac{\mu_C}{kT} - 1\right) \quad (3)$$

$$GR_Q = G_{CQ} - (R_{CQ} + R_{NC})\left(exp\frac{\mu_C - \mu_Q}{kT} - 1\right) \quad (4)$$

$$GR_Q = G_{QV} - R_{QV}\left(exp\frac{\mu_Q}{kT} - 1\right) - R_{NV}\left(1 - exp\frac{-\mu_Q}{kT}\right) \qquad (5)$$

Here $G_{ji}$ and $R_{ji}$ are coefficients related to absorption and recombination, respectively; $n_B$ is the density of electrons in the buffer; $\mu_Q = F_Q - F_V$ is the split of quasi-Fermi level of holes in QDs; $\varepsilon_B$ is the height of blocking barrier; $\varepsilon_{B0}$ is the height of blocking barrier when the cell is in the thermal equilibrium, $V = G = \mu_C = \mu_Q = 0$; and $\mu_C$ is the split of quasi-Fermi level of electrons accumulated in the conduction band of QD absorber,

$$exp\frac{\mu_C}{kT} = \frac{n_B}{n_{B0}} \qquad (6)$$

Due to the continuity of electron flow through QDs, $\mu_Q$ balances the net electron transitions from the electronic states confined in QD with transitions into those states. Equalizing of Equations (4) with (5) yields

$$exp\frac{\mu_Q}{kT} = \frac{G_{QV} - G_{CQ} - R_{CQ} + R_{QV} - R_{NV} - R_{NC}}{R_{QV}}\left[\frac{1}{2} \pm \sqrt{\frac{1}{4} + \frac{\frac{1}{R_{QV}}\left[R_{NV} + (R_{CQ} + R_{NC})\frac{n_B}{n_{B0}}\right]}{\left(1 + \frac{G_{QV} - G_{CQ} - R_{NV} - R_{CQ} - R_{NC}}{R_{QV}}\right)^2}}\right] \qquad (7)$$

Here sign " + " should be used when $G_{QV} - G_{CQ} - R_{CQ} - R_{NV} - R_{NC} + R_{QV} \geq 0$, and sign " − " should be use when $G_{QV} - G_{CQ} - R_{CQ} - R_{NV} - R_{NC} + R_{QV} \leq 0$.

*2.4. Coefficients*

Assuming absorption of all incoming photons in each of $[\varepsilon_i, \varepsilon_j]$ spectral ranges, one can use the principle of detailed balance to reduce the rate of electron transitions related to solar photon absorption $G_{ji}$ to integrals [21]

$$G_{ji} = \frac{2Xe}{h^3c^2A_S}\int_{\varepsilon_i}^{\varepsilon_j}\frac{\varepsilon^2 d\varepsilon}{exp(\varepsilon/kT_S) - 1} \qquad (8)$$

where $X$ is the concentration of solar light, and $A_S = 4.6 \times 10^4$ is the geometrical factor related to the angle that Earth is seen from Sun.

On the other hand electron-hole radiative recombination yields photon emission from the cell. Coefficients $R_{ji}$ related to such emission at temperature $T$ can be written in a form similar to Equation (8) [21],

$$R_{ji} = \frac{2e}{h^3c^2}\int_{\varepsilon_i}^{\varepsilon_j}\frac{exp(-\mu/kT)\varepsilon^2 d\varepsilon}{exp[(\varepsilon-\mu)/kT] - 1} \qquad (9)$$

where $\mu$ is the split for the relevant quasi-Fermi level.

Also the coefficient $R_{NC}$ related to electron-hole non-radiative inter-band generation-recombination in QDs can be written as [5]

$$R_{NC} = e(L_A n_B/\tau_n)exp(\varepsilon_B/kT) \qquad (10)$$

where $\tau_n$ is the lifetime of non-radiative inter-band recombination through QDs, and $L_A$ is the thickness of the QD absorber.

Assuming the same quasi-Fermi level for mobile holes in absorber and buffer, one can reduce the coefficient $R_{NV}$ related to non-radiative relaxation of holes from the mobile into the confined electronic states in QDs to [5]

$$R_{NV} = e(N_Q L \Omega N_B / \tau_{ph}) exp(-\varepsilon_B / kT) \qquad (11)$$

where $N_Q$ is the density of QDs in absorber; $\Omega$ is the average volume of QDs; $\tau_{ph}$ is the lifetime of intra-band relaxation in QDs, which occurs usually due to inelastic scattering of holes on optical phonons, $\tau_{ph} \approx 1 ps$.

*2.5. Diffusion equation*

Both concentrated sunlight and bias voltage break the balance of carriers in p-n-junction. This leads to accumulation of mobile electrons in the conduction band of QD absorber and splits the Fermi level as shown in Figure 3.

Negative charge of accumulated electrons reduces the height of blocking barrier from $\varepsilon_{B0}$ to $\varepsilon_B$, and increases the electron density in the buffer $n_{B0}$ to $n_B$. The latter is a parameter $n_B$ of the Poisson Equation (1) regulating $\varepsilon_B$ value of the blocking barrier.

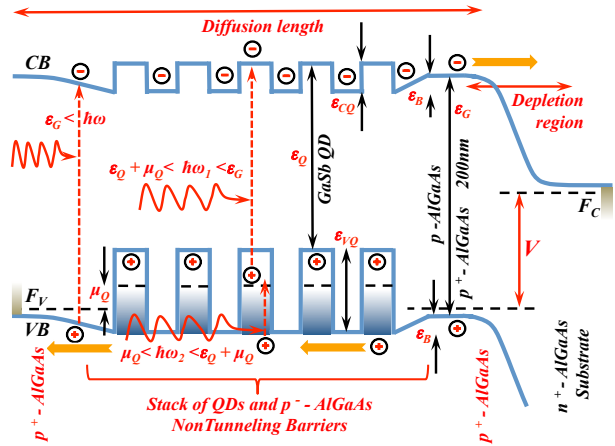

**Figure 3.** Modification of the energy band bending shown in Figure 2 by concentrated sunlight and bias $V$. Accumulation of negative charge of $n(x)$ electrons in the conduction band of GaSb/GaAs type-II QD absorber moves up the band edges in the absorber. The same charge reduces $\varepsilon_{B0}$ blocking barrier that spatially separates the QD absorber from the depletion region of p-n-junction as shown in Figure 2.

Accumulated electrons also escape over the blocking barrier into the n-doped region. Equation $D_n \partial^2 n(x)/\partial x^2 = -GR/L_A$ of electron diffusion in the conduction band describes this escape from the QD absorber and determines the parameter $n_B$ of Equation (1) as a function of the bias voltage, $V$, and the concentration of sunlight, $X$. Assuming a small gradient of the quasi-Fermi level $\mu_C(x)$ for mobile electrons accumulated in the conduction band in the QD absorber, we can rewrite the diffusion equation as

$$D_n \frac{\partial^2 n(x)}{\partial x^2} = \frac{\frac{n(x)}{n_{B0}}\left(R_{CV} + (R_{CQ}+R_{NC})exp\frac{-\mu_Q}{kT}\right)exp\frac{-\varepsilon_B}{kT} - R_{CV} - R_{CQ} - R_{NC} - G_{CV} - G_{CQ}}{L_A} \qquad (12)$$

where $D_n$ is the electron diffusion coefficient, $L_A$ is the thickness of QD absorber, and $n(x)$ is the density of accumulated electrons, $n(x) = n_{B0} exp[(\mu_C(x) + \varepsilon_B)/kT]$.

*2.6. Accumulation and currents*

Integrating of diffusion Equation (12) over the absorber yields the rate of photoelectron injection from the absorber into the buffer, we obtain

$$J_A = G_{CV} + G_{CQ} + (R_{CV} + R_{CQ} + R_{NC})\left(1 - \frac{n_B}{n_{B0}}\frac{\left(R_{CV} + (R_{CQ}+R_{NC})exp\frac{-\mu_Q}{kT}\right)}{(R_{CV}+R_{CQ}+R_{NC})}\right) \qquad (13)$$

On the other hand, electron diffusion through the buffer that is $L_B$ thick yields an injection flow over the blocking barrier

$$J_B = e D_n (n_B - n_{B0} exp(eV/kT))/L_B \qquad (14)$$

Equalizing of Equations (13) and (14) for the continuity of electron currents through the interface yields equations that determine the parameter $n_B$ and the photocurrent $J_B$

$$n_B = \frac{(G_{CV}+G_{CQ})\frac{L_B}{eD_n}}{1+\frac{R_{CV}L_B}{eD_n n_{B0}}\left(1+\frac{R_{CQ}+R_{NC}}{R_{CV}}exp\frac{-\mu_Q}{kT}\right)} + \frac{exp\frac{eV}{kT}+\frac{R_{CV}L_B}{eD_n n_{B0}}\left(1+\frac{R_{CQ}+R_{NC}}{R_{CV}}\right)}{1+\frac{R_{CV}L_B}{eD_n n_{B0}}\left(1+\frac{R_{CQ}+R_{NC}}{R_{CV}}exp\frac{-\mu_Q}{kT}\right)} n_{B0} \qquad (15)$$

$$J_B = \frac{G_{CV}+G_{CQ}+R_{CV}\left(1-exp\frac{eV}{kT}\right)+(R_{CQ}+R_{NC})\left(1-exp\frac{eV-\mu_Q}{kT}\right)}{1+\frac{R_{CV}L_B}{eD_n n_{B0}}\left(1+\frac{R_{CQ}+R_{NC}}{R_{CV}}exp\frac{-\mu_Q}{kT}\right)} \qquad (16)$$

Inserting $J_B = 0$ into Equation (16) yields the open circuit voltage, $V_{oc}$, of the cell

$$exp\frac{eV_{oc}}{kT} = \frac{G_{CV}+G_{CQ}+R_{CV}+R_{CQ}+R_{NC}}{R_{CV}+(R_{CQ}+R_{NC})exp\left(-\frac{\mu_{Qoc}}{kT}\right)} \qquad (17)$$

where $\mu_{Qoc}$ is the quasi-Fermi level given by Equation (7) where $V = V_{oc}$.

Inserting $V = 0$ into Equation (16) yields the short-circuit current, $J_{sc}$, of the cell

$$J_{sc} = \frac{G_{CV}+G_{CQ}+(R_{CQ}+R_{NC})\left(1-exp\frac{-\mu_{Qsc}}{kT}\right)}{1+\frac{R_{CV}L_B}{eD_n n_{B0}}\left(1+\frac{R_{CQ}+R_{NC}}{R_{CV}}exp\frac{-\mu_{Qsc}}{kT}\right)} \qquad (18)$$

where $\mu_{Qsc}$ is the quasi-Fermi level given by Equation (7) where $V = 0$.

In the dark, $X = 0$, Equation (16) yields the dark current $J_{dk}$ of proposed solar cell,

$$J_{dk} = \frac{R_{CV}\left(1-exp\frac{eV}{kT}\right)+(R_{CQ}+R_{NC})\left(1-exp\frac{eV-\mu_{Qd}}{kT}\right)}{1+\frac{R_{CV}L_B}{eD_n n_{B0}}\left(1+\frac{R_{CQ}+R_{NC}}{R_{CV}}exp\frac{-\mu_{Qd}}{kT}\right)} \qquad (19)$$

where $\mu_{Qd}$ is the quasi-Fermi level given by Equation (7) where $X = 0$.

Equalizing Equation (16) with the sum of Equations (18) and (19) yields the equation of congruency

$$\frac{G_{CV}+G_{CQ}+R_{CV}\left(1-exp\frac{e(V+\delta V)}{kT}\right)+(R_{CQ}+R_{NC})\left(1-exp\frac{e(V+\delta V)-\mu_Q}{kT}\right)}{1+\frac{R_{CV}L_B}{eD_n n_{B0}}\left(1+\frac{R_{CQ}+R_{NC}}{R_{CV}}exp\frac{-\mu_Q}{kT}\right)} - $$
$$\frac{G_{CV}+G_{CQ}+(R_{CQ}+R_{NC})\left(1-exp\frac{-\mu_{Qsc}}{kT}\right)}{1+\frac{R_{CV}L_B}{eD_n n_{B0}}\left(1+\frac{R_{CQ}+R_{NC}}{R_{CV}}exp\frac{-\mu_{Qsc}}{kT}\right)} - \frac{R_{CV}\left(1-exp\frac{eV}{kT}\right)+(R_{CQ}+R_{NC})\left(1-exp\frac{eV-\mu_{Qd}}{kT}\right)}{1+\frac{R_{CV}L_B}{eD_n n_{B0}}\left(1+\frac{R_{CQ}+R_{NC}}{R_{CV}}exp\frac{-\mu_{Qd}}{kT}\right)} = 0 \qquad (20)$$

This Equation (20) determines $\delta V$ divergence of the dark current curve from the photocurrent curve at a given photocurrent in the cell.

## 3. Results and discussion

### 3.1. Key features of the proposed model

#### 3.1.1. Spatial separation from the depletion region

The major key feature of the proposed solar cell is spatial separation of QD absorber from the depletion region as shown in Figure 1. We already showed that the built-in field of depletion region might facilitate recombination through IB and QD states even in ideal p-n-junction [22]. We also pointed out that spatial separation of those IB and QD electronic states from the depletion region adds more flexibility to the cell design in sense of improving the device characteristics [15]. In particular, such separation may eliminate additional dark current related to a leakage through IB and QD electronic states.

The next novel feature is involvement of the type-II QDs instead of commonly used type-I QDs in solar cells. Due to the type-II misalignment of energy bands, type-II QDs have $\varepsilon_{CQ}$ offset shown in Figures 2 and 3 that creates local barriers shaped like "anti-dots" in the conduction band of type-II QD absorber. These "anti-dot" barriers spatially separate mobile electrons of conduction band from holes confined in type-II QDs as shown in Figure 4.

Let us recall that type-I QDs capture both electrons and holes. Therefore, spatial separation that eliminates electron tunneling into and from type-II QDs also reduces recombination coefficient $\gamma_R$ of type-II QDs in comparison to that of type-I QDs. Reduction of $\gamma_R$ may be sufficient for even moderate concentration of sunlight to facilitate two-photon absorption of sub-bandgap photons and generation of the additional photocurrent in QD solar cells due to such absorption [23].

We assume incoming sunlight is concentrated so much that it is able to change distribution of both mobile and confined holes in the valence band of QD absorber as well as distribution of mobile photoelectrons in the conduction band. This change of distribution splits the Fermi level into quasi-Fermi levels $F_C$, $F_V$ and $F_Q$. Figure 3 shows that $F_C$ relates to mobile electrons in the conduction band, $F_V$ relates to mobile holes in the valence band, and $F_Q = F_V + \mu_Q$ relates to holes confined in QDs.

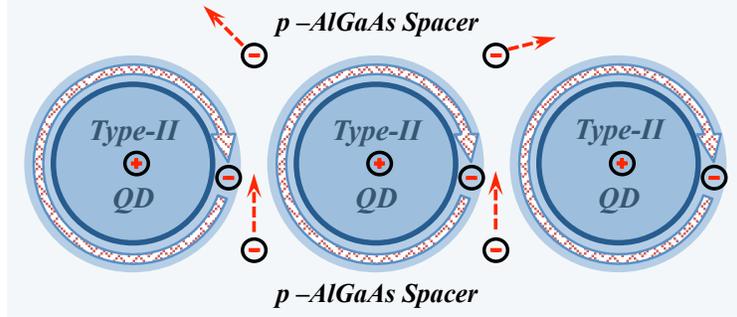

**Figure 4.** A schematic picture demonstrating motion of electrons in the conduction band of GaSb/GaAs type-II QD absorber spatially separated from the depletion region. Electrons pass between GaSb QDs surrounded with p-doped $Al_xGa_{1-x}As$ spacer layers.

We assume the same $F_V$ quasi-Fermi level for mobile holes in absorber, cap, and buffer layers. For this to take place, mobile holes should pass through the QD absorber so quickly that they come to equilibrium with all mobile holes of p-doped region. As a consequence of the same $F_V$ quasi-Fermi level, the $p^+$-doped $Al_xGa_{1-x}As$ buffer acquires a positive charge to balance diffusion of mobile holes into the buffer.

Another important point is that we consider type-II QD absorber spatially separated but still within the diffusion length from the depletion region. When a bias $V > 0$ is applied in the dark, $G = 0$, it facilitates intensive electron injection from the n-doped region into the conduction band of the QD absorber. This injection sets the same quasi-Fermi level for conduction band electrons in QD absorber and n-doped region of p-n-junction. Therefore, $\mu_C = eV$ and $\varepsilon_B = \varepsilon_{B0} - eV$. However, illumination with concentrated sunlight, $G \gg 0$, breaks this balance of the quasi-Fermi levels. It pushes the quasi-Fermi level up in the absorber in comparison with that in the n-doped region, which makes $\varepsilon_B \leq \varepsilon_{B0} - eV$.

Let us recall that two-photon absorption transfers mobile photoelectrons, first, to the top of $\varepsilon_{CQ}$ "anti-dot" barriers in the conduction band of type-II QDs shown in Figure 4. These photoelectrons have few $10^7 cm/s$ thermal velocity, therfore promptly escape from the few nm height QDs into the $Al_xGa_{1-x}As$ spacers in about $10fs$. They lose $\varepsilon_{CQ}$ excess energy in collisions with optical phonons and relax to the conduction band edge in the spacers in about $1ps$. Since relaxation is irreversible process, the relaxed electrons face $\varepsilon_{CQ}$ local "anti-dot" barriers in the conduction band. It is important that these "anti-dot" barriers do not limit electron diffusion in the conduction band while spatially separate them from holes confined in the type-II QDs as shown in Figure 4.

It should be noted that in addition graded $Al_xGa_{1-x}As$ spacers of QDs can be used in the absorber. By pulling photoelectrons, the built-in field of such spacers enforces drift-diffusion in the conduction band so much that photoelectrons may pass through $1\mu m$ thick QD absorber in $0.1ns$, which is less than inter-band recombination lifetime,

$1 ns - 10 ns$ [24].

*3.1.2. Non-radiative recombination*

In this work we add into consideration two non-radiative electron transitions associated with QDs, in addition to the radiative electron transitions discussed in the frame of the detailed balance principle in [4]. The first of those transitions is given by Equations (10) and represents the inter-band recombination of the conduction band electrons with holes confined in QDs. The second transition is given by Equations (11) and represents the intra-band relaxation of mobile holes in QDs from mobile into confined electronic states due to inelastic scattering of holes on optical phonons in QDs.

It should be noted that spatial separation of mobile electrons from confined holes in type-II QD absorber slows down both radiative and non-radiative inter-band recombination at QDs. Experiments with strain-compensated GaSb/GaAs type-II QDs shown that inter-band recombination at QDs may be reduced so much that electron lifetime in the type-II QD absorber becomes as long as it in the bulk GaAs, $\tau_n = 10\ ns$ [24].

*3.1.3. Blocking barrier $\varepsilon_B$*

Figure 1 displays the proposed structure of GaAs solar cell. The QD absorber is a stack of alternating strain-compensated GaSb type-II QDs and GaAs spacers. A thin p$^+$-doped Al$_x$Ga$_{1-x}$As buffer layer spatially separates such GaSb/GaAs type-II QD absorber from the depletion region of p-n-junction. The absorber is about un-doped and covered with p$^+$-doped Al$_x$Ga$_{1-x}$As cap layer. Together buffer and cap sandwich the absorber as shown in Figure 1.

For equalizing the Fermi level across the structure, mobile holes inject into the QD absorber from both p$^+$-doped layers. Such injection known as the modulation doping of semiconductor heterojunctions [25] determines the energy band bending and the density of mobile holes in proposed QD absorber. Positive charge of injected holes lowers by $\varepsilon_{B0}$ the conduction and valence band edges in QD absorber relative to that in the p$^+$-doped buffer layer as shown in Figure 2. The bending of energy bands also builds up the energy barrier, $\varepsilon_B$, at the absorber/buffer interface, which blocks diffusion of conduction band electrons from the absorber into the n-doped region of p-n-junction.

The bending of energy bands increases the density of mobile electrons and reduces the density of mobile holes $exp(\varepsilon_B/kT)$ times in the conduction and valence bands of QD absorber with respect to their densities in the buffer layer.

When the cell is not biased, $V = 0$, and it is in the dark, $G = 0$, the electron density in the buffer, $n_B$, gets the lowest value of $n_B = n_{B0}$. For this lowest value of parameter $n_B$, solution of Poisson Equation (1) yields the highest value of the blocking barrier $\varepsilon_B = \varepsilon_{B0}$ shown in Figure 2.

Illumination with concentrated sunlight, $G \sim X > 0$, facilitates both single- and two-photon absorption in the QD absorber. The red dashed arrows shown in Figure 3 denote electron transfers associated to both absorption mechanisms. Concentrated sunlight transfers a huge number of mobile electrons from the valence band into the conduction band of the QD absorber. Accumulation of mobile electrons in the conduction band of QD absorber is the clue to understanding of photovoltaic performance of proposed solar cells. While the blocking barrier limits escape of mobile electrons from the absorber, the negative charge of accumulated electrons reduces the blocking barrier from $\varepsilon_{B0}$ to a smaller $\varepsilon_B$ value. For instance, as shown in Figure 5 with the dark red dots, already concentration of about 300-sun may reduce the blocking barrier $\varepsilon_B$ to the thermal energy, $26 meV$, of mobile carriers. In fact, such barrier is very small to limit electron diffusion and, hence, to influence on photovoltaic performance of solar cells.

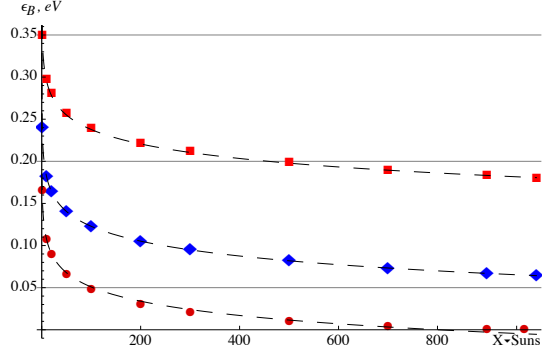

**Figure 5.** The height of screening barrier $\varepsilon_B$ as a function of sunlight concentration $X$ for GaAs solar cell with spatially separated GaSb/GaAs type-II QD absorber. The dark red circles relate to the radiative limit. The red squares and blue diamonds relate to the non-radiative intra-band relaxation of holes in GaSb/GaAs type-II QDs when the relaxation lifetimes are $1 ps$ and $100 ps$, respectively. Dashed curves of $a + b \times ln(X)$ approximations and solid lines are shown for guidance.

Therefore, accumulation of mobile electrons facilitates their injection from the absorber through the buffer into the n-doped region of p-n-junction. A small gradient of their density in the conduction band of QD absorber maintains electron diffusion that balances electron injection through the interface for the continuity of electron currents. Equations (12) and (13) describe this diffusion.

*3.2. Calculation of parameters and currents*

*3.2.1. Extraction of parameters $\varepsilon_B$, $n_B$, $\mu_Q$, $\mu_C$ and photocurrent $J_B$ as a function of bias voltage $V$ and concentration $X$*

Solution of Equations (1) to (15) enables calculation of photovoltaic characteristics of proposed QD solar cell. We started from Poisson equation (1), where we used parameters $n_B = n_{B0}$ and $G = 0$, for calculating the blocking barrier $\varepsilon_B = \varepsilon_{B0}$ in the dark. Then we inserted $\mu_C$, $\mu_Q$, and coefficients $G_{ij}$, $R_{ij}$, $R_{Nj}$ given by Equations (6) to (11) into Equation (15), which made it a transcendent equation. We solved this transcendent equation with respect to parameter $n_B$ as a function of $\varepsilon_B$. In the next step we inserted the obtained solution for the parameter $n_B$ as a function of $\varepsilon_B$ into Equation (1). This allowed us to find the blocking barrier $\varepsilon_B < \varepsilon_{B0}$ and the increased electron density $n_B$ as a function of applied bias voltage $V$ and concentration $X$.

Then we substituted parameters $\varepsilon_B$ and $n_B$ with their values obtained as functions of bias voltage $V$ and concentration $X$ in Equations (6), (7) and (14). Such substitution yielded also the quasi-Fermi levels $\mu_Q$ and $\mu_C$ and the photocurrent $J_B$ as functions of bias voltage $V$ and sunlight concentration $X$.

*3.2.2. Photocurrent*

Let us denote with $L_D$ the electron diffusion length in the p-doped region of the proposed solar cell. Electrons that pass through the buffer contribute in the photocurrent. Equation (16) describes this photocurrent. The denominator of Equations (16) includes recombination losses related to $R_{CV}$, $R_{CQ}$ and $R_{NC}$ recombination intensities, where $R_{CV} \approx eD_n n_{B0}/L_D$. Contribution of these losses depends on the ratios of the recombination intensities to the diffusion flow through the buffer, $eD_n n_{B0}/L_B$. All values are at the thermal equilibrium. Since the ratio $R_{CV} L_B/(eD_n n_{B0}) \approx L_B/L_D \ll 1$, Equations (15) - (20) can be simplified. In particular, the photocurrent given by Equation (16) reduces to

$$J_B = \frac{G_{CV} + G_{CQ} + R_{CV}\left(1 - exp\frac{eV}{kT}\right) + (R_{CQ} + R_{NC})\left(1 - exp\frac{eV - \mu_Q}{kT}\right)}{1 + \frac{(R_{CQ} + R_{NC})L_B}{eD_n n_{B0} exp(\mu_Q/kT)}} \qquad (21)$$

We assume that the reference solar cell was made along with the proposed single junction solar cell but without QDs. In such a cell $R_{CQ} = R_{NC} = G_{CQ} = 0$. Inserting these values into Equation (21) yields the photocurrent $J_{Ref}$ of the reference solar cell,

$$J_{Ref} = G_{CV} + R_{CV}\left(1 - exp\frac{eV}{kT}\right) \qquad (22)$$

*3.2.3. Additional photocurrent*

By comparing Equations (21) and (22) of photocurrents generated in the proposed and reference solar cells, we can understand the impact of concentrated sunlight on performance of the proposed solar cell.

Illumination yields $\mu_Q < 0$ when the sub-bandgap photons of solar spectrum remove more electrons than holes from the QD confined electronic states. Such illumination facilitates recombination through QDs. If $(R_{CQ} + R_{NCQ})L_B \gg eD_n n_{B0} exp(\mu_Q/kT)$, Equation (21) has large denominator that suppresses the photocurrent. The latter becomes small, $J_B \ll G_{SCV} + G_{SCQ}$, even at $V = 0$ because only a few photoelectrons generated in the conduction band of QD absorber are able to contribute into the photocurrent.

At $J = 0$, the numerator of Equation (21) yields the open circuit voltage of the cell. It has a smaller value that that of the reference cell because $\mu_Q < 0$. A reduced open circuit voltage is often observed in conventional solar cells and interpreted as an arrest of the quasi-Fermi level by uncontrollable defects and impurities embedded in these cells [26].

It should be noted that illumination yields $\mu_Q > 0$ in the proposed solar cell when sub-bandgap photons of solar spectrum remove more holes than electrons from the electronic states confined in QDs. Although these QDs as artificial impurities embedded in the absorber would facilitate recombination, Equation (21) unveils that there is an exception.

QDs do not facilitate electron-hole recombination when $\mu_Q > 0$ is large enough so that

$$eD_n n_{B0} exp(\mu_Q/kT) \gg (R_{CQ} + R_{NC})L_B \qquad (23)$$

This condition reduces Equation (21) of the photocurrent to,

$$J_B = G_{CV} + R_{CV}\left(1 - exp\frac{eV}{kT}\right) + G_{CQ} + (R_{CQ} + R_{NC})\left(1 - exp\frac{eV - \mu_Q}{kT}\right) \qquad (24)$$

The first two terms in the right side of Equation (24) reduce to photocurrent $J_{Ref}$ of the reference solar cell given by Equation (22); the last two terms represent the additional photocurrent related to the absorption in QDs. Because of $R_{CV} \approx eD_n n_{B0}/L_D$, condition (23), and $L_B/L_D \ll 1$, Equation (24) yields a little bit smaller value of the open circuit voltage in QD solar cells than that in the reference solar cell as shown in Figure 6.

Let us focus on Equations (23) and (24). Importantly, under the condition described by Equation (23), QDs reduce the open circuit voltage less than they facilitate generation of the additional photocurrent as shown in Figure 6.

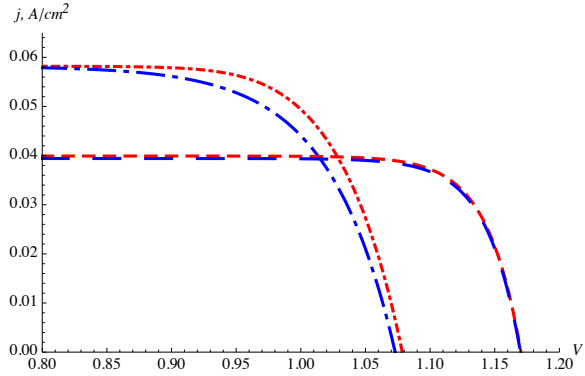

**Figure 6.** Current-voltage characteristics of GaAs solar cells. Photocurrents are calculated for 1-sun concentration. The GaSb/GaAs type-II QD absorber is spatially separated from the depletion region of p-n-junction. The short-dashed-dot red curve relates to the photocurrent of QD solar cell. The short-dashed red curve relates to the photocurrent of the reference cell assuming condition (23). The long-dashed-dot blue curve relates to the dark current of QD solar cell. The long-dashed blue curve relates to the dark current of the reference cell. Dark currents are superimposed on the relevant photocurrents. The break of congruency is clearly seen in $I - V$ curves of QD solar cell above $0.8V$ bias.

Figure 6 displays the photo and dark currents as functions of the bias voltage $V$. The dashed red curve refers to $J_{Ref}$ of the reference cell given by Equation (22) while the solid red curve refers the photocurrent $J_B$ given by Equation (24). The blue curves refer to the relevant dark currents. The shape of these $I - V$ curves looks like they represent ideal solar cells where absorption of sub-band gap photons simply increases the photocurrent while almost not affecting the open circuit voltage.

*3.2.4. Dark current*

Let us denote $\mu_{Qd}$ the quasi-Fermi level $\mu_Q$ given by Equation (7) of proposed solar cell in the dark, where $X = 0$ and $G_{CV} = G_{CQ} = G_{QV} = 0$, and substitute these values for $\mu_Q$ and the relevant $G_{ji}$ coefficients in Equation (16). Such substitution yields Equation (19) of the dark current of proposed QD solar cell. Since usually $eD_n n_{B0} \gg L_B R_{CV}$ and there are no QDs in the reference cell, $R_{CQ} = R_{QV} = R_{NC} = 0$, Equation (19) yields the same dark current, $J_{dRef} = R_{CV}(1 - exp(eV/kT))$, for both reference and ideal p-n-junction solar cells.

Figure 7 displays the dark current curve of proposed solar cell in semi-logarithmic scale. There are two linear regions of the curve on the plot. The break of slope is clearly seen in the plot at $0.55V$ bias. The ideality factors $n = 1$ and $n = 2$ fit to the slopes of dark current curve in the left and right sides from the break, respectively. Competition of electron-hole recombination in the QD absorber with electron injection from the n-doped region through the buffer into the QD absorber determines the shape of the dark current curve. Injection through the buffer limits the dark current when the bias is less than $0.55V$ while recombination in the QD absorber limits the dark current when the bias rises over $0.55V$.

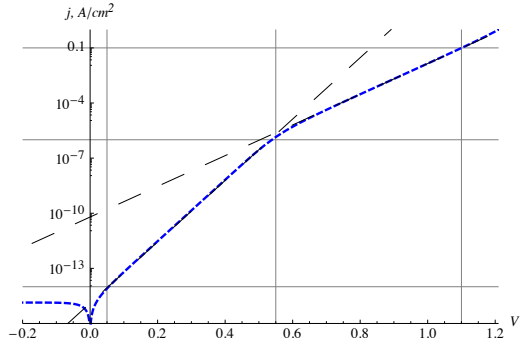

**Figure 7.** Dark current-voltage characteristics of GaAs solar cell with GaSb/GaAs type-II QD absorber spatially separated from the depletion region of p-n-junction. The break of slope is clearly seen on the small-dashed curve at $0.55V$ bias. The slope of this curve corresponds to the ideality factor $n = 1$ and $n = 2$ in the left and right sides from the break, respectively. The long-dashed and solid lines are shown for guidance.

It shows that the right-hand side of Equation (16) includes Equation (19) of the dark current. In the dark the terms proportional to $exp(eV/kT)$ in the right hand side of Equation (16) persist. These terms describe electron injection into the QD absorber from the n-doped region of cell when bias is $V$. We can compare the shape of both photo and dark currents by superimposing their curves at $V = 0$ and studying the divergence of those curves. Equation (20) gives this divergence $\delta V$ at a given current such that $\delta V = 0$ at $V = 0$. These curves are congruent, $\delta V \equiv 0$, for instance, for the ideal p-n-junction without QDs, $R_{CQ} = R_{QV} = R_{NC} = G_{CQ} = G_{QV} = 0$, as shown in Figure 6.

Involvement of QDs breaks congruency of the superimposed photo and dark current curves of proposed solar cell. Figure 6 displays the divergence of those curves. Such non-congruency is often seen for photo and dark current curves in experiments with thin film solar cells [27]. Usually the observed non-congruency is referred to accumulation of charge in the depletion region [28]. However, the proposed solar cell accumulates negatively charged electrons within the conduction band of QD absorber. This region is spatially separated from the depletion region.

*3.2.5. Quasi-Fermi level $\mu_Q$ of confined holes*

Condition (23) and Equation (24) of photocurrent are insensitive to the blocking barrier $\varepsilon_B$ but exponentially depend on the quasi-Fermi level $\mu_Q$ of confined holes. This quasi-Fermi level regulates accumulation of charge in QDs by balancing photoelectron generation and their recombination in QDs. Equation (7) describes occupation of confined electronic states in QDs and determines numerical value of $\mu_Q$. This value is a function of bias voltage $V$ and sunlight concentration $X$ as shown in Figure 8. The red curve displays the occupancy $\mu_Q$ for $X = 1$ concentration of sunlight and the blue curve displays that in the dark, $X = 0$. Both curves are insensitive to small bias $V$ but arise linearly when the bias is large. The break of the red and blue curves occurs at $V_L = 0.8V$ and $V_L = 0.5V$, respectively. As shown in Figure 8, both lines coincide at high bias voltages and their slope is $\partial \mu_Q / \partial eV = 0.5$ above the breaking points.

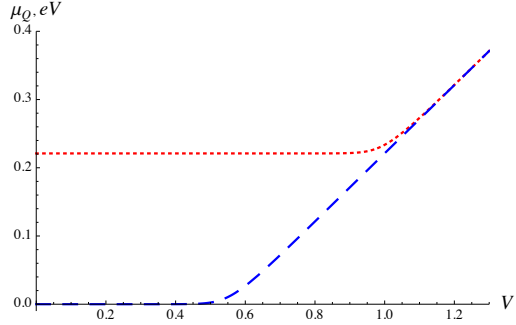

**Figure 8.** Quasi-Fermi level $\mu_Q$ of holes confined in GaSb/GaAs type-II QDs of the QD absorber as a function of bias $V$. The red-dotted and blue-dashed curves represent calculation for $X = 1$ sun concentration of sunlight and in the dark, $X = 0$, respectively.

Such behavior of the quasi-Fermi level results from Equation (7). Let`s take into account that $\mu_C = eV$ and that Equation (6) determines the ratio $n_B/n_{B0}$. It is easy to see that when $V > V_L$, Equation (7) yields the linear dependence of $\mu_Q$ on $V$ with the slope of $\partial \mu_Q / \partial eV = 0.5$,

$$\mu_Q = 0.5eV - kT \times ln[R_{QV}/(R_{CQ} + R_{NC})] \tag{25}$$

where $V_L$ is the breaking point,

$$V_L = \frac{kT}{e} \times ln \frac{(G_{QV} - G_{CQ} + R_{QV} - R_{NV} - R_{CQ} - R_{NC})^2}{R_{QV}(R_{CQ} + R_{NC})} \tag{26}$$

Since both $G_{QV} \sim X$ and $G_{CQ} \sim X$, this breaking point $V_L$ depends on concentration of sunlight. The higher concentration of sunlight, the higher bias $V_L$ is needed for the break shown in Figure 8.

On the other hand when $V < V_L$, Equation (7) yields a constant $\mu_{QL}$ with regard to $V$,

$$\mu_{QL} = kT \times ln[1 + (G_{QV} - G_{CQ} - R_{CQ} - R_{NV} - R_{NC})/R_{QV}] \quad (27)$$

Numerical value of this constant $\mu_{QL}$ depends on concentration $X$ of sunlight because both $G_{QV} \sim X$ and $G_{CQ} \sim X$. In the dark, $G_{CQ} = G_{QV} = 0$, and Equation (27) yields $\mu_{QL} \approx 0$. Switching of illumination from the dark to one-sun, $X = 1$, increases $\mu_{QL}$ by $0.22 eV$ and $V_L$ by $0.45 V$ as shown in Figure 8. Concentration in hundreds suns may increase $\mu_{QL}$ up to $0.4 eV$ as shown in Figure 9.

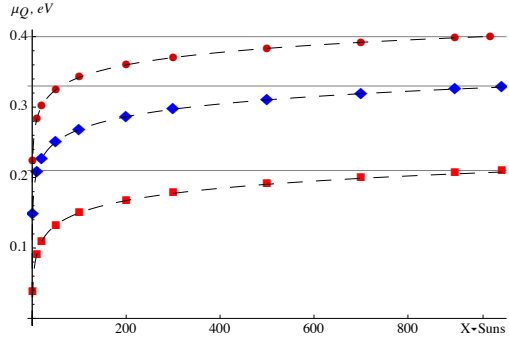

**Figure 9.** Quasi-Fermi level $\mu_Q$ of confined holes as a function of sunlight concentration $X$ for GaAs solar cell with spatially separated GaSb/GaAs type-II QD absorber. The dark red circles relate to the radiative limit. The red squares and blue diamonds relate to the non-radiative intra-band relaxation of holes in GaSb/GaAs type-II QDs when the relaxation lifetimes are $1ps$ and $100ps$, respectively. Dashed curves of $a + b \times ln(X)$ approximations and solid lines are for guidance.

It should be noted here that both inter-band and intra-band non-radiative electron transitions through QDs reduce the quasi-Fermi level $\mu_Q$ and increase the blocking barrier $\varepsilon_B$. For instance, non-radiative relaxation of holes in QDs reduces $\mu_Q$ by $0.07 eV$ and $0.18 eV$ as shown in Figure 9 while it increases $\varepsilon_B$ by $0.07 eV$ and $0.14 eV$ as shown in Figure 5 when relaxation lifetime is *100ps* and *1ps*, respectively.

Equation (27) unveils an important feature of the proposed solar cell. Concentration of sunlight increases $\mu_Q$ and may bring the cell to operation under condition (23). Recall that condition (23) leads to about ideal characteristics shown in Figure 6.

*3.3. Photovoltaic characteristics*

*3.3.1. Open circuit voltage and short circuit current*

The open circuit voltage $V_{oc}$ is given by Equation (17). For ideal p-n-junction without QDs, $R_{CQ} = R_{QV} = R_{NC} = 0$, Equation (17) yields $V_{ocR} = (kT/e) \times ln(G_{CV}/R_{CV})$. Since $G_{CV} \sim G_{CQ} \sim X$, Equation (17) yields a logarithmical increase of $V_{oc}$ with concentration of sunlight as shown in Figure 10.

Involvement of QDs activates both radiative and non-radiative recombination in QD absorber. Too intensive recombination, $R_{CQ} + R_{NC} > R_{CV} exp(\mu_{Qoc}/kT)$, reduces the open circuit voltage. For instance, intra-band relaxation of mobile holes into confined states in QD absorber may reduce $V_{oc}$ by $0.25 eV$ as shown in Figure 10.

On the other hand, condition (23) eliminates recombination through QDs. Therefore, Equations (22) and (24) yield about the same value of the open circuit voltage for the reference $V_{ocR}$ and the proposed $V_{oc}$ solar cells. The difference of these voltages is only $0.1V$ as shown in Figure 6, where intersection of the red curves of photocurrents with the $V-$axis at $J = 0$ determines the open circuit voltages.

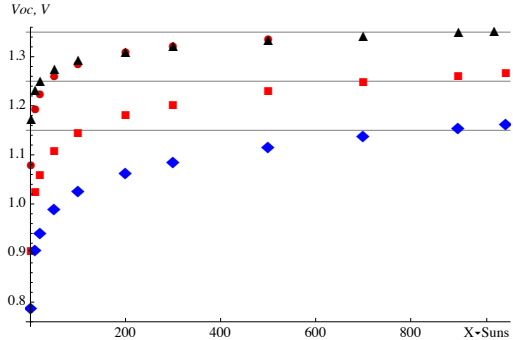

**Figure 10.** Open circuit voltage $V_{oc}$ as a function of sunlight concentration $X$ for GaAs solar cell with GaSb/GaAs type-II QD absorber spatially separated from the depletion region. The dark red circles relate to the radiative limit. The red squares and blue diamonds indicate the non-radiative intra-band relaxation of holes in GaSb/GaAs type-II QDs when the relaxation lifetimes are $1ps$ and $100ps$, respectively. Solid lines are for guidance. The black triangles relate to Shockley-Queisser limit of the reference GaAs solar cell.

The short circuit current is given by Equation (18). For ideal p-n-junction without QDs, $R_{CQ} = R_{QV} = R_{NC} = 0$, Equation (18) yields $J_{shR} = G_{CV}$. Also Equations (22) and (24) yield $J_{shR} = G_{CV}$ and $J_{shQ} = G_{CV} + G_{CQ}$ at $V = 0$ for the short circuit currents of the reference $J_{shR}$ and the proposed $J_{shQ}$ solar cells, respectively. Such values of the short circuit currents point out complete collection of all mobile photoelectrons generated in the reference cell and also in the conduction band of the QD absorber under the condition (23), respectively.

There is no explicit dependence of Equation (23) on the blocking barrier $\varepsilon_B$. Then the condition (23) leads to complete collecting of generated electrons even when the blocking barrier is large, $\varepsilon_B > kT$. Noteworthy this example unveils that in some cases blocking barriers do not harm photoelectron collection.

*3.3.2. Fill-factor FF*

Figure 11 displays the fill-factor of proposed QD solar cell as a function of sunlight concentration. The fill-factor *FF* of ideal QD solar cell (dark red dots) is smaller than that of the ideal reference GaAs solar cell (black dots). However, while the latter is about fixed at 91% and insensitive to sun-concentration, the fill-factor of proposed cell (dark red dots) grows to that value along with concentration of sunlight. For concentration of about 300-sun, the fill-factor turns to saturation at the same 91%.

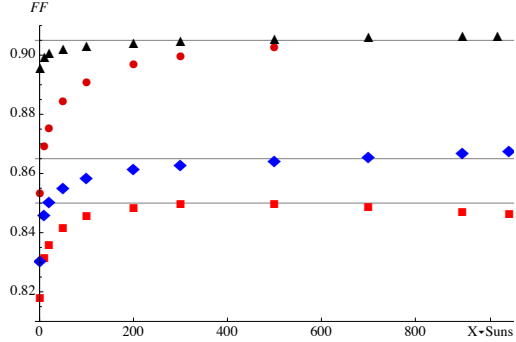

**Figure 11.** Fill-Factor *FF* as a function of sunlight concentration *X* for GaAs solar cell with GaSb/GaAs type-II QD absorber spatially separated from the depletion region. The dark red circles relate to the radiative limit. The red squares and blue diamonds relate to the non-radiative intra-band relaxation of holes in GaSb/GaAs type-II QDs when the relaxation lifetimes are $1ps$ and $100ps$, respectively. Solid lines are for guidance. The black triangles relate to Shockley-Queisser limit of the reference GaAs solar cell.

Figure 11 also shows that non-radiative relaxation of holes in QD absorber reduces the fill-factor of proposed QD solar cell. The red and the blue dots in Figure 11 are calculated for *1ps* and *100ps* photoelectron relaxation lifetimes in QDs, respectively. The faster the intra-band relaxation, the lower the saturation value that fill-factor gets. Like the open circuit voltage shown in Figure 10, the same 300-sun concentration turns also the "non-radiative" fill-factor *FF* to saturation.

### 3.3.3. Conversion efficiency η

Figure 12 displays the dependence of conversion efficiency $\eta$ on concentration *X* of sunlight for proposed and reference solar cells. Concentration of sunlight is needed for achieving of higher performance. Black dots denote the Shockley-Queisser limit for the reference GaAs solar cell while dark red dots denote the radiative recombination limits for the proposed GaSb/GaAs QD IB solar cell. Concentration of sunlight increases the conversion efficiency. Approximations $a + b \times ln(X)$ given by the solid lines in Figure 12 fit well to the calculated efficiencies. The fit mirrors the similar dependence of the open circuit voltage $V_{oc}$ on concentration *X* shown in Figure 10.

When concentration *X* of sunlight is strong enough to facilitate two-photon absorption in QDs it also pushes up the conversion efficiency $\eta$ of GaSb/GaAs QD IB cell along with generation of additional photocurrent. The efficiency of the proposed solar cell may reach the Luque-Marti limit, which is 51.6% for $X = 1000$-sun concentration [8], and the difference in efficiencies of the reference and proposed solar cells may rise up to 17% as shown in Figure 12. Such enhancement of efficiency becomes possible because accumulation of charge in the QD absorber about eliminates the blocking barrier $\varepsilon_B$ as shown in Figure 5. This barrier becomes so small that cannot already limit photoelectron escape over the barrier through the buffer layer from the QD absorber into the depletion region. Therefore, further concentration of sunlight has a little effect.

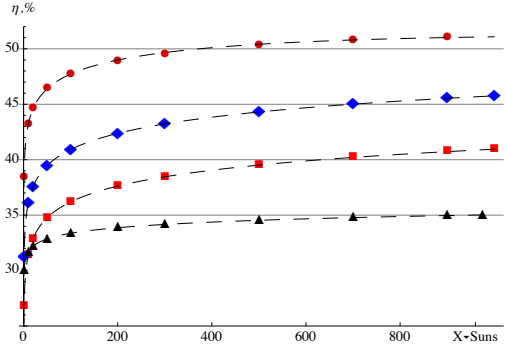

**Figure 12.** Conversion efficiency $\eta$ as a function of sunlight concentration $X$ for GaAs solar cell with GaSb/GaAs type-II QD absorber spatially separated from the depletion region. The dark red circles relate to the radiative limit. The red squares and blue diamonds relate to the non-radiative intra-band relaxation of holes in GaSb/GaAs type-II QDs when the relaxation lifetimes are $1ps$ and $100ps$, respectively. The black triangles relate to Shockley-Queisser limit of the reference GaAs solar cell. Dashed curves of $a + b \times ln(X)$ approximations and solid lines show the grid.

Photo-induced reduction of the blocking barrier $\varepsilon_B$ is highly sensitive to non-radiative electron transitions in QD absorber. Inter-band recombination of mobile photoelectrons with confined holes, $R_{NC}$, and intra-band relaxation of holes in QDs, $R_{NV}$, restrain such reduction of the blocking barrier and, hence, reduce the efficiency. For instance, the red and blue dots in Figure 12 display such reduction of the conversion efficiency when non-radiative intra-band relaxation lifetime of holes in QDs is $\tau_{ph} = 1ps$ and $\tau_{ph} = 100ps$, respectively, and non-radiative inter-band recombination of mobile electrons with holes in QDs is $\tau_C = 10ns$. Experiments showed that $\tau_C = 10ns$ [24] while $\tau_{ph} = 1ps$ or may be longer [29]. It should be noted that though these non-radiative relaxation and recombination degrade the performance of proposed solar cell, its conversion efficiency is still above the Shockley-Queisser limit by 5% to 10%.

## 4. Conclusions

QD solar cells have a potential to display superior conversion efficiency due to generation of additional photocurrent by sub-bandgap photons of solar spectra, thereby motivating a search of novel structures able to raise the conversion efficiency of solar cells above the 37% of Shockley-Queisser limit. Practical realization of these concepts encountered technological problems connected with formation of high-density QD arrays with low defect densities in solar cells since, like artificial atoms, QDs may act as fast recombination centers facilitating generation of additional dark current [5,8]. However, recently essential progress has been achieved in technology of formation of strain-compensated InAs QD arrays in GaAs solar cells [12, 13].

Another important development in the field was recognition [5, 14] that type-II QDs have an attractive potential for solar cell application. There are also relevant developments with indirect band gap Si/Ge type-II QDs [15, 16].

We emphasize a critical issue that so far, to the best of our knowledge, all models and experiments on QD solar cells studied QDs located in the depletion region, which is a "default" location of QDs in optoelectronic applications. Our study found, however, that there is no benefit from such location of QDs in solar cells. Moreover, we showed that solar cells might benefit from spatial separation of type-II QDs from the depletion region.

Looking for a new practical structure for IB QD solar cell application, we focused on a stack of strain-compensated GaSb/GaAs type-II QDs and application of such stack as the QD absorber of sub-band gap photons [5]. We proposed a novel structure with GaSb/GaAs type-II QD absorber embedded in the p-doped region of ideal solar cell, but spatially separated from the depletion region. We developed the model of such QD solar cell and used the detailed balance principle along with Poisson and continuity equations for calculating of the energy band bending along with the photocurrent and the dark current, and the conversion efficiency of the cell. Our model takes into account both single-photon and double-photon absorption as well as non-radiative processes in QDs, like intra-band relaxation of holes in QDs and inter-band electron-hole recombination in type-II QDs, embedded in the p-doped region of otherwise ideal solar cell.

We found that concentration of sunlight improves performance of the proposed solar cell and predicted that concentration from 1-sun to 500-sun raises the efficiency from 30% to 50%. We showed that accumulation of charge in QD absorber is the clue to understanding of superior performance. An attractive feature of the proposed solar cell is that QDs do not reduce the open circuit voltage but facilitate generation of the additional photocurrent to the extent that photovoltaic characteristics reach ideal IB solar cell while the efficiency meets the Luque-Marti limit. Noteworthy, although non-radiative processes like relaxation in QDs and recombination through QDs degrade photovoltaic characteristics of the proposed solar cell, its conversion efficiency is still predicted to be above the Shockley-Queisser limit by 5% to 10%.

Our study demonstrated important steps toward producing practical QD solar cells that benefit from additional photocurrent generated by sub-band gap photons. A US patent on design and manufacturing of the proposed device has been applied for [30].

**Acknowledgments**

A. Kechiantz and A. Afanasev acknowledge support from The George Washington University and Dominion Resources, Inc. J.-L. Lazzari acknowledge support from CNRS.